\documentclass[aps,prl,twocolumn,groupedaddress,showpacs,showkeys]{revtex4}

\usepackage{graphicx}
\usepackage{bm}
\usepackage{amsmath}

\begin{document}

\title{Bright gap solitons of atoms with repulsive interaction}

\author{B. Eiermann$^1$, Th. Anker$^1$,
M. Albiez$^1$, M. Taglieber$^1$, P. Treutlein$^{2}$, K.-P.
Marzlin$^3$, and M.K. Oberthaler$^1$} \affiliation{
$^{1}$Kirchhoff Institut f\"ur Physik, Universit{\"a}t Heidelberg,
Im Neuenheimer Feld 227, 69120 Heidelberg, Germany\\
$^{2}$Max-Planck-Institut f\"{u}r Quantenoptik und Sektion Physik
der Ludwig-Maximilians-Universit{\"a}t, Schellingstr.4, 80799
M\"{u}nchen, Germany \\ $^3$Quantum Information Science Group
Department of Physics and Astronomy 2500 University Drive NW
Calgary, Alberta T2N 1N4 Canada}
\date{\today}

\begin{abstract}
We report on the first experimental observation of bright
matter-wave solitons for $^87$Rb atoms with repulsive atom-atom
interaction. This counter intuitive situation arises inside a weak
periodic potential, where anomalous dispersion can be realized at
the Brillouin zone boundary. If the coherent atomic wavepacket is
prepared at the corresponding band edge a bright soliton is formed
inside the gap. The strength of our system is the precise control
of preparation and real time manipulation, allowing the systematic
investigation of gap solitons.
\end{abstract}


\maketitle

Non-spreading localized wave packets \cite{russel} - bright
solitons - are a paradigm of nonlinear wave dynamics and are
encountered in many different fields, such as physics, biology,
oceanography, and telecommunication. Solitons form if nonlinear
dynamics compensates the spreading due to linear dispersion.

For atomic matter waves, bright solitons have been demonstrated
where the linear spreading due to vacuum dispersion is compensated
by the attractive interaction between atoms
\cite{bright_solitons}. For repulsive atom-atom interaction dark
solitons have also been observed experimentally
\cite{dark_solitons}.

In this letter we report on the experimental observation of a
different class of solitons, which only exist in periodic
potentials - bright gap solitons. For weak periodic potentials
formation of atomic gap solitons has been predicted
\cite{gapsolitons_bec} while discrete solitons
\cite{discretesolitons} should be observable in the case of deep
periodic potentials. These phenomena are well known in the field
of nonlinear photon optics where the nonlinear propagation
properties in periodic refractive index structures have been
studied \cite{opticalgapsolitons}. In our experiments with
interacting atoms a new level of experimental control can be
achieved allowing for the realization of standing gap solitons for
repulsive atom-atom interaction corresponding to a
self-defocussing medium. It also opens up the way to study driven
solitons \cite{townssoliton} and the realization of two- and three
dimensional discrete solitons \cite{2D_3Dsolitons}.

In our experiment we investigate the evolution of a Bose-Einstein
condensate in a quasi one-dimensional waveguide with a weak
periodic potential superimposed in the direction of the waveguide.
In the limit of weak atom-atom interaction the presence of the
periodic potential leads to a modification of the linear
propagation i.e. dispersion \cite{modification_linear}. It has
been demonstrated that anomalous dispersion can be realized with
this system \cite{dispersionmanagement} which is the prerequisite
for the realization of bright gap solitons for repulsive atom-atom
interaction.

\begin{figure}[b]
\includegraphics[width=0.83\linewidth]{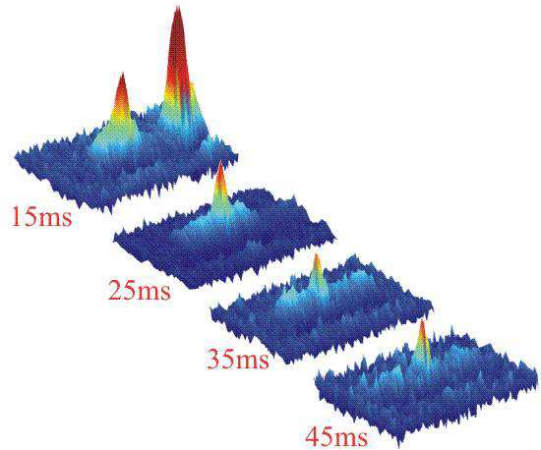}
\caption{\label{fig:1}Observation of bright atomic gap solitons.
The atomic density in the negative mass regime deduced from
absorption images (430$\mu$m x 125$\mu$m) averaged over 4
realizations is shown for different propagation times. After
approximately 25ms a small peak is formed which does neither
change in shape nor in amplitude. Excessive atoms are radiated and
disperse over time. After 45ms only the soliton with $\sim$250
atoms has sufficient density to be clearly observable. The second
peak at 15ms shows the atoms which have been removed by Bragg
scattering to generate an initial coherent wave packet consisting
of $\sim$900 atoms.}
\end{figure}

Our experimental observations are shown in figure \ref{fig:1} and
clearly reveal that after a propagation time of 25ms a
non-spreading wave packet is formed. The observed behavior
exhibits the qualitative features of  bright gap soliton formation
such as: (a) during soliton formation excessive atoms are radiated
and spread out over time (b) solitons do not change their shape
and atom number during propagation (c) gap solitons do not move.

\begin{figure}[ht]
\includegraphics[width=0.9\linewidth]{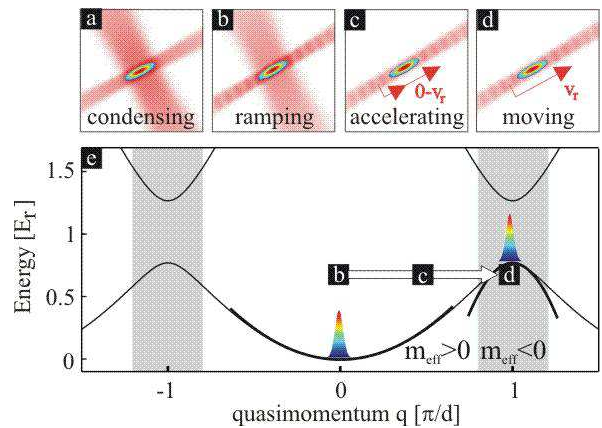}
\caption{\label{fig:2} Realization of coherent atomic wavepackets
with negative effective mass utilizing periodic potentials. (a)
top view of the crossed dipole trap geometry used for
Bose-Einstein condensation. (b) a periodic potential is ramped up
while the atoms are still trapped in the crossed dipole trap
realizing the atomic ensemble at $q_c=0$. (c,d) the atoms are
released into the one-dimensional waveguide and subsequently the
periodic potential is accelerated to the recoil velocity $v_r=
h/\lambda m$. This prepares the atomic wavepacket at the band edge
of the lowest band. (e) normal and anomalous dispersion regime in
a periodic potential. The single preparation steps are indicated.
The shown band structure is calculated for the experimentally
employed potential modulation depth of V=0.7 $E_r$ leading to
$m_{\mathrm{eff}}/m=-0.1$ at the band edge.}
\end{figure}

The coherent matter-wave packets are generated with $^{87}$Rb
Bose-Einstein condensates (figure \ref{fig:2}a). The atoms are
initially precooled in a magnetic TOP trap using the standard
technique of forced evaporation leading to a phase space density
of $\sim$0.03. The atomic ensemble is subsequently adiabatically
transferred into a crossed light beam dipole trap
($\lambda$=1064nm, 1/$e^2$ waist 60 $\mu$m, 500mW per beam) where
further forced evaporation is achieved by lowering the light
intensity in the trapping light beams. With this approach we can
generate pure condensates with typically $3\times10^4$ atoms. By
further lowering the light intensity we can reliably produce
coherent wave packets of 3000 atoms. For the successful
demonstration of gap solitons further reduction of the atom number
is necessary. For that purpose we employ a Bragg pulse in the trap
allowing the preparation of coherent wave packets of $\sim$ 900
atoms.

After the preparation of the coherent wave packet a periodic
dipole potential realized with a far off-resonant standing light
wave of wavelength $\lambda=783$nm (figure \ref{fig:2}b), is
adiabatically ramped up. As indicated in figure~\ref{fig:2} this
procedure prepares the atomic ensemble in the normal dispersion
regime at quasimomentum $q=0$. The dispersion relation for an atom
moving in a weak periodic potential exhibits a band structure as a
function of quasimomentum  $q$ known from the dispersion relation
of electrons in crystals \cite{Ashcroftenglish76} (see figure
\ref{fig:2}e). Anomalous dispersion, characterized by a negative
effective mass $m_{\mathrm{eff}}<0$, can be achieved if the mean
quasimomentum of the atomic ensemble is shifted to the Brillouin
zone boundary corresponding to $q=\pi/d$. This is accomplished by
switching off one dipole trap beam, releasing the atomic cloud
into the one-dimensional horizontal waveguide (Fig.~\ref{fig:2}c)
with transverse and longitudinal trapping frequencies
$\omega_\perp =2\pi \times 85\mathrm{Hz}$ and $w_{||}= 2\pi \times
0.5\mathrm{Hz}$. Subsequently the atomic ensemble is prepared at
quasimomentum $q=\pi/d$ by accelerating the periodic potential to
the recoil velocity $v_r= h/m\lambda$. The acceleration within
1.3ms is adiabatic, therefore excitations to the upper bands are
negligible \cite{Landau_tunneling}. It is important to note that
the strength of dispersion and correspondingly the absolute value
of $m_{\mathrm{eff}}$ is under full experimental control, since it
scales with the modulation depth of the periodic potential.

For weak periodic potentials the full wavefunction of the
condensate is well described by $\Psi(x,t)=A(x,t) u_0^{q_c}(x)
\exp(iq_cx)$, where $u_0^{q_c}(x)\exp(iq_cx)$ represents the Bloch
state in the lowest band $n=0$ at the corresponding central
quasimomentum $q_c$. Within the approximation of constant
effective mass, the dynamics of the envelope $A(x,t)$ is governed,
by a one-dimensional nonlinear Schr\"odinger equation
\cite{Steel98}
\begin{equation}
i \hbar \frac{\partial}{\partial t} A(x,t) =
\left(-\frac{\hbar^2}{2m_{\mathrm{eff}}}\frac{\partial^2}{\partial
x^2} + g_{1d} |A(x,t)|^2\right) A(x,t)\nonumber
\end{equation}
with $g_{1d}=2\hbar a \omega_{\perp} \alpha_{nl}$ where
$\alpha_{nl}$ is a renormalization factor due to the presence of
the periodic potential ($\alpha_{nl}$ =1.5 for $q=\pi/d$ in the
limit of weak periodic potentials \cite{Steel98}), and a is the
scattering length. The stationary solution for $q_c=\pi/d$ is
given by
\begin{equation}
A(x,t) = \sqrt{N/2x_0} \mathrm{sech}(x/x_0) e^{i \hbar t/2
m_{\mathrm{eff}} x_0^2}, \nonumber
\end{equation}
where $x_0$ is the soliton width and $m_{\mathrm{eff}}$ is the
effective mass at the band edge.

The quantitative features of bright solitons can be understood
through comparison of the characteristic energies for dispersion
and atom-atom interaction. The linear spreading is characterized
by the energy $E_D = \hbar^2/2m_{\mathrm{eff}}x_0^2$. The
atom-atom interaction is given by $E_{nl}= g_{1d} |A(x=0,t)|^2$.
Equating both energies, leads to the total number of atoms
constituting the soliton
\begin{equation}
    N= \frac{\hbar}{\alpha_{nl} a \omega_{\perp} m_{\mathrm{eff}} x_0}.
    \label{number}
\end{equation}
A characteristic time scale of solitonic propagation due to the
phase evolution can also be identified. In analogy to light optics
the soliton period is given by $T_S = \pi m_{\mathrm{eff}} x_0^2/2
\hbar$. Solitonic propagation can be confirmed experimentally if
the wave packet does not broaden over time periods much longer
than $T_S$.

Our experimental results in figure \ref{fig:1} show the evolution
of a matter wave soliton in the negative mass regime for different
propagation times. The reproducible formation of a single soliton
is observed if the initial wave packet is close to the soliton
condition, i.e. a well defined atom number for a given spatial
width. Our experimental setup allows the reliable production of
condensates which contain not less than 3000 atoms with a spatial
size of $\sim 2.5 \mu$m (rms) resulting from the longitudinal
trapping frequency of the crossed dipole trap $\sim 40\times
2\pi$Hz. For this atom number no gap solitons have been observed.
Therefore we remove 70\% of the atoms by Bragg scattering leading
to an inital wavepacket with 900(300) atoms. As shown in figure
\ref{fig:1}, the Bragg scattered atoms are still visible after
15ms and move out of the imaged region for longer observation
times. The soliton can clearly be distinguished from the
background after 25ms, corresponding to 3 soliton periods. This is
consistent with the typical formation time scale of few soliton
periods given in nonlinear optics text books \cite{Agrawal01}.
After 45ms of propagation, the density of the radiated atoms drops
below the level of detection and thus a pure soliton remains,
which has been observed for up to 65ms. In order to understand the
background we numerically integrated the nonpolynomial nonlinear
Schr\"odinger equation \cite{salasnich}. The calculation reveals
that the non-quadratic dispersion relation in a periodic potential
leads to an initial radiation of atoms. However the absolute
number of atoms in the observed background is much higher than the
prediction of the employed effective one-dimensional model.
Therefore we conclude that transverse excitations have to be taken
into account to get quantitative agreement. This fact still has to
be investigated in more detail.

\begin{figure}[t]
\includegraphics[width=0.95\linewidth]{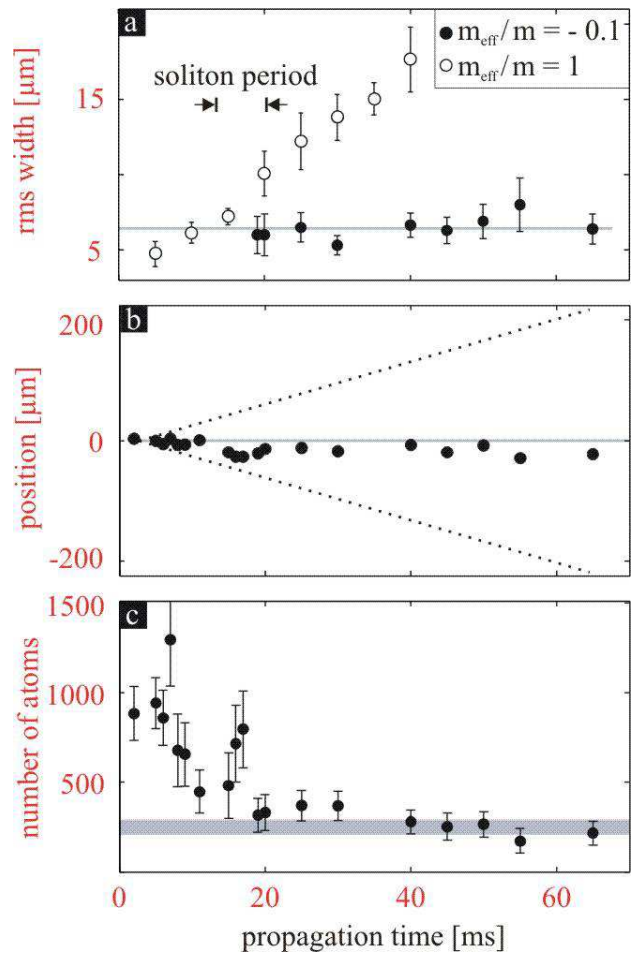}
\caption{\label{fig:3}Characteristic features of the observed gap
soliton. (a) Comparison of expansion in the positive and negative
effective mass regime for 300 atoms. While the soliton does not
disperse at all over a time of 65ms, corresponding to more than 8
soliton periods (solid circles), a wave packet in the normal mass
regime expands significantly (open circles). Each point represents
the result of a single realization. (b) shows the position of the
soliton in the frame of the periodic potential and reveals that a
standing gap soliton has been realized. The dotted lines indicate
the positions that correspond to maximum and minimum group
velocity in the lowest band.  (c) number of atoms in the central
peak. The initial atom numbers exhibit large shot to shot
fluctuations, which are reduced during the soliton formation. The
predicted relation between the number of atoms and the soliton
width (eq. 1) is indicated by the horizontal bar in graph c using
the width deduced as shown in graph (a). Note that this comparison
has been done without free parameter and all contributing
parameters are measured independently.}
\end{figure}

\begin{figure}[b]
\includegraphics[width=0.95\linewidth]{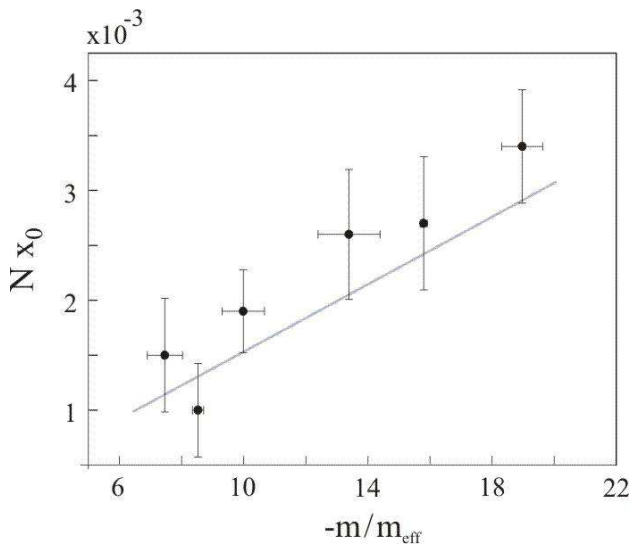}
\caption{\label{fig:4}Scaling properties of an atomic gap soliton.
The effective mass was varied experimentally by changing the
periodic potential depth. The scaling predicted by (eq.1)  is
represented by the solid line and is in excellent agreement with
our experimental observations. The errorbars represent the
variation of the scaling parameter for different realizations.  }
\end{figure}

In the following we will discuss the experimental facts shown in
figure \ref{fig:3} and figure \ref{fig:4} confirming the
successful realization of atomic bright gap solitons.

In figure \ref{fig:3}a we compare the spreading of wave packets in
the normal and anomalous regime which reveals the expected
dramatic difference in wave packet dynamics. The solid circles
represent the width of the gap soliton  for
$m_{\mathrm{eff}}/m=-0.1$, which does not change significantly
over time. In this regime, the wave packet does not spread for
more than 8 soliton periods ($T_S=7.7$ms). We deduce a soliton
width of $x_0=6(1) \mu m$ ($x_{rms}=4.5\mu$m) from the absorption
images where the measured rms width shown in figure \ref{fig:3}a
is deconvolved with the optical resolution of $3.8 \mu$m (rms).
Since our experimental setup allows to switch from solitonic to
dispersive behavior by turning the periodic potential on and off,
we can directly compare the solitonic evolution to the expected
spreading in the normal dispersion regime. The open circles
represent the expansion of a coherent matter wave packet with
300(100) atoms in the normal mass regime $m_{\mathrm{eff}}/m=1$.

The preparation at the band edge implies that the group velocity
of the soliton vanishes. This is confirmed in figure \ref{fig:3}b,
where the relative position of the soliton with respect to the
standing light wave is shown. The maximum group velocity of the
lowest band is indicated with the dotted lines. In the experiment
care has to be taken to align the optical dipole trap
perpendicular to the gravitational acceleration within
200$\mu$rads. Otherwise the solitons are accelerated in the
direction opposite to the gravitational force revealing their
negative mass characteristic.

The calculated number of atoms (eq.\ref{number}) is indicated by
the horizontal bar in figure \ref{fig:3}c. The width of the bar
represents the expectation within our measurement uncertainties.
The observed relation between atom number and width,
characteristic for a bright soliton, is in excellent agreement
with the simple theoretical prediction without any free parameter.

As an additional check for soliton formation, we determine the
product of atom number and soliton width as a function of the
effective mass which is varied by adjusting the modulation depth
of the periodic potential. Figure \ref{fig:4} shows the range of
effective masses, for which solitons have been observed. For
smaller values of $|m_{\mathrm{eff}}|$, corresponding to smaller
potential depths, Landau-Zener Tunneling does not allow a clean
preparation in the negative mass regime, while for larger values
the initial number of atoms differs too much from the soliton
condition. The observed product of atom number and wave packet
width after 40ms of propagation are shown in figure \ref{fig:4}
and confirm the behaviour expected from eq.1. Additionally, our
experimental findings reveal that the change of the scaling
parameter $Nx_0$ in figure \ref{fig:4} is dominated by the change
in the atom number, while the soliton width only exhibits a weak
dependence on the effective mass.

The demonstration of standing atomic gap solitons confirms that
Bose condensed atoms combined with a periodic potential allow the
precise control of dispersion and nonlinearity. Thus our setup
serves as a versatile new model system for nonlinear wave
dynamics. Our experiments show that atomic gap solitons can be
created in a reproducible manner. This is an essential
prerequisite for the study of soliton collisions. The experiment
can be realized by preparing two spatially separated wavepackets
at the band edge and applying an expulsive potential. Ultimately,
atom number squeezed states can be engineered with atomic solitons
by implementing schemes analog to those developed for photon
number squeezing in light optics \cite{Friberg}. This is
interesting from a fundamental point of view and may also have
impact on precision atom interferometry experiments.

We wish to thank J. Mlynek for his generous support, Y. Kivshar,
E. Ostrovskaya,  A. Sizmann and B. Brezger for many stimulating
discussions. We thank O. Vogelsang and D. Weise for their donation
of Ti:Sapphire light. This work was supported by Deutsche
Forschungsgemeinschaft, Emmy Noether Program, and by the European
Union, Contract No. HPRN-CT-2000-00125.

\end{document}